\documentclass[aps,pre,preprint,showpacs]{revtex4}
\usepackage{epsf}
\usepackage{amsmath,amssymb}

\begin{document}

\title {Dephasing and delay time fluctuations in the chaotic scattering
of a quantum particle\\
weakly coupled to a complicated background}

\date{\today}

\author{Valentin V. Sokolov} \affiliation{Budker Institute
of Nuclear Physics, Novosibirsk, Russia; Center for
Nonlinear and Complex Systems, Universita degli Studi
dell'Insubria, Como, Italy}

\begin{abstract}
Effect of a complicated many-body environment is analyzed on the chaotic motion of a
quantum particle in a mesoscopic ballistic structure. The dephasing and absorption
phenomena are treated on the same footing in the framework of a schematic microscopic
model. The single-particle doorway resonance states excited in the structure via an
external channel are damped not only because of the escape onto such channels but also
due to ulterior population of the long-lived background states. The transmission through
the structure is presented as an incoherent sum of the flow formed by the interfering
damped doorway resonances and the retarded flow of the particles reemitted by the
environment. The resulting internal damping as well as the dephasing rate are uniquely
expressed in terms of the spreading width which controls the coupling to the background.
The formation of the long-lived fine-structure resonances strongly enhances delay time
fluctuations thus broadening the delay time distribution.
\end{abstract}


\maketitle

1. Analog experiments with open electromagnetic microwave cavities
\cite{Stoeckmann1990,Doron1990,Alt1995,Schaefer2003} on the one hand and extensive study
of the electron transport through ballistic microstructures \cite{Huibers1998} on the
other have drawn during last decade much attention to peculiarities of chaotic wave
interference in open billiard-like set-ups. It is well recognized by now that the
statistical approach \cite{Verbaarschot1985,Beenakker1997,Alhassid2000} based on the
random matrix theory (RMT) provides a reliable basis for describing the universal
fluctuations characteristic of such an interference which, in particular, manifests
itself in the single-particle resonance chaotic scattering and transport phenomena. At
the same time, experiments with ballistic quantum dots reveal persisting up to very low
temperatures appreciable deviations from the predictions of RMT which  indicate some
loss of the quantum coherence.

Two different methods of accounting for the dephasing effect have been suggested which
give different results. In the B\"uttiker's voltage-probe model \cite{Buettiker1986} an
subsidiary randomizing scatterer is introduced with $M_{\phi}$ channels each with a
sticking coefficient $P_{\phi}$. Even assuming all these channels to be equivalent we
are still left with two independent parameters. This results in an ambiguity since
quantities of physical interest (e.g. the conductance distribution) depend, generally,
on $M_{\phi}$ and $P_{\phi}$ separately whereas the dephasing phenomenon is controlled
by the unique parameter: the dephasing time $\tau_\phi$ which is fixed by their product.
On the other hand, only the strength of a uniform imaginary potential governs the
Efetov's model \cite{Efetov1995} but, implying absorption, this model suffers a loss of
unitarity. The deficiencies and differences of the two models
\cite{McCLer1996,Brouwer1997ii} have been analyzed, in particular in
\cite{Brouwer1997ii}. A prescription was suggested how to get rid of uncertainties, and
simultaneously, to accord the models by considering the limit
$M_{\phi}\!\rightarrow\!\infty$, $P_{\phi}\!\rightarrow 0$ at fixed product
$M_{\phi}\!P_{\phi}\!\equiv\!\Gamma_{\phi}$ in the first model and by compulsory
restoration of the unitarity in the second. The construction proposed infers a
complicated internal structure of the probe. Otherwise, the assumed limit could hardly
be physically justified. If so, the typical time $\tau_p$ spent by the scattering
particle inside the probe, being proportional to its mean spectral density, forms,
generally, a new time scale different from the dephasing time. A question thus arises on
the total time delay during transport through a ballistic mesoscopic structure in
presence of a complicated background.

2. Below we propose a model of dephasing and absorption phenomena, which from the very
beginning does not suffer any ambiguity. We consider the environment as a complicated
many-body system with a very dense energy spectrum. The evolution of the extended
system: the moving inside the structure particle and the background coupled to each
other with an interaction $V$, is described by means of an enlarged non-Hermitian
Hamiltonian matrix $\mbox {\boldmath ${\cal H}$}$ of order $N\!=\!N^{(s)}\!+\!N^{(e)}$
\begin{equation}\label{Htot}
\mbox {\boldmath ${\cal H}$}=
\left(\begin{array}{cc} {\cal H}^{(s)} & V^{\dagger} \\
        V  & {\cal H}^{(e)}  \end{array}\right)
\end{equation}

The two blocks ${\cal
H}^{(r)}\!=\!H^{(r)}\!-\!\frac{i}{2}A^{(r)}{A^{(r)}}^{\dag},r\!=\!s, e$ ($N^{(e)}\gg
N^{(s)}$) along the main diagonal represent the non-Hermitian effective Hamiltonians of
the bare single-particle open system and the environment respectively. The Hermitian
matrices $H^{(r)}$ describe their closed counterparts (the corresponding mean level
spacings satisfy the condition $D(r\!=\!s)\gg d(r\!=\!e)$) when the matrices $A^{(r)}$
are built of the amplitudes connecting the internal and channel states. The two
(generally unstable) subsystems are mixed by the off-diagonal $N^{(e)}\times N^{(s)}$
coupling matrix $V$. When uncoupled, these subsystems have no common decay channels so
that the coupling $V$ is purely Hermitian. This means that the background states have no
direct access to the observable outer channels and attain it only due to the mixing with
the single-particle "doorway" states inside the dot.

The total $M\times M$ scattering matrix ${\bf S}(E)\!=\!{\bf I}-i{\bf A}^{\dag}
\left(E-\mbox {\boldmath ${\cal H}$}\right)^{-1}{\bf A}$ is unitary so far as all the
$M\!=\!M^{(s)}+ M^{(e)}$ open channels, observable $(s)$ as well as hidden $(e)$, are
taken into account. However, only transitions between $M^{(s)}$ outer channels are
accessible for direct observations. They are described by the $M^{(s)}\times M^{(s)}$
submatrix $S(E)\!=\!I-i{\cal T}(E)$ where
\begin{equation}\label{T}
{\cal T}(E)={A^{(s)}}^{\dagger}{\cal
G}^{(s)}_{\textsc{d}}(E)A^{(s)}
\end{equation}
and the matrix
\begin{equation}\label{GD}
{\cal G}^{(s)}_{\textsc{d}}(E)=\frac{1}{E-{\cal H}^{(s)}-
\Sigma^{(s)}(E)}
\end{equation}
is the upper left $N^{(s)}\times N^{(s)}$ block of the resolvent $\mbox {\boldmath
${\cal G}$}(E)\!=\! (E-$\mbox {\boldmath ${\cal H}$}$)^{-1}$. The subscript $D$ stands
in (\ref{GD}) for "doorway". The self-energy matrix
\begin{equation}\label{Sig}
\Sigma^{(s)}(E)=V^{\dagger}\frac{1}{E-{\cal
H}^{(e)}}V\equiv V^{\dagger}{\cal G}_0^{(e)}(E)V\,.
\end{equation}
includes all virtual transitions between the bare doorway and environment resonances.
This matrix is not Hermitian as long as the background states are not stable. Therefore
the submatrix $S$ is not, generally, unitary. The flow of outgoing particles through
outer channels reduces because of inelastic processes in the background thus implicating
absorption.

We further assume that the coupling matrix elements $V_{\mu m}$, where the indices $\mu$
and $m$ span the background's and the system's Hilbert spaces respectively, are random
Gaussian variables, $\langle V_{\mu m}\rangle$=$0$; $\langle V_{\mu m}V^{*}_{\nu
n}\rangle$=$\delta_{\mu\nu}\delta_{mn} \frac{1}{2}\Gamma_s\frac{d}{\pi}$. Here
$\Gamma_s= 2\pi\langle |V|^2\rangle/d$ is the spreading width
\cite{BohrMot1969,Mahaux1969,Sokolov1997}. The inequality $\langle
|V|^2\rangle$$\gg$$d^2$ is implied so that the interaction $V$, though weak, is not weak
enough for perturbation theory to be valid. Retaining the original notations also for
the quantities averaged over the interaction $V$, we find in the main $1/N^{(e)}$
approximation
\begin{equation}\label{avG}
{\cal G}^{(s)}_{\textsc{d}}(E)=\left[E-\frac{1}{2}\Gamma_s
g^{(e)}(E)- {\cal H}^{(s)}\right]^{-1}\;.
\end{equation}
where $g^{(e)}(E)$=$\frac{d}{\pi}$$\rm Tr$${\cal G}_0^{(e)}(E)$. The resonance spectrum
$\{{\cal E}_{\alpha}\!=\!E_{\alpha}\!-\!\frac{i}{2}\Gamma_{\alpha}\}$ is defined in such
an approximation from $N^{(s)}$ similar equations
\begin{equation}\label{Rs}
{\cal E}-\frac{1}{2}\Gamma_s g^{(e)}({\cal E})- {\cal
H}^{(s)}=0
\end{equation}
originated each from one of the bare single-particle doorway resonances.

In what follows we analyze in detail temporal aspects of the scattering where the
influence of the background shows up in the most full and transparent way. In
particular, a straightforward calculation gives for the Smith delay-time submatrix
$Q=-iS^{\dag}dS/dE$ in the $M^{(s)}$-dimensional subspace of the observable channels
$Q\!=\!{\bf {b^{(s)}}}^{\dag}{\bf b^{(s)}}
\!=\!{b^{(s)}}^{\dag}b^{(s)}+{b^{(e)}}^{\dag}b^{(e)} \!=\!Q^{(s)} + Q^{(e)}$ where the
matrix amplitudes
\begin{equation}\label{bb}
{\bf b^{(s)}}(E)=\mbox {\boldmath ${\cal G}$}(E)A^{(s)}\,;
\quad b^{(s)}(E)={\cal G}^{(s)}_{\textsc{d}}(E)A^{(s)}\,;
\quad b^{(e)}(E)={\cal G}_0^{(e)}(E)Vb^{(s)}(E)
\end{equation}
have dimensions $N\times M^{(s)}$, $N^{(s)}\times M^{(s)}$ and $N^{(e)}\times M^{(s)}$
respectively. The two contributions $Q^{(s,e)}$ correspond to the modified due to the
interaction with the background time delay within the dot and delay because of the
virtual transitions into the background. After averaging over the interaction $V$ we
arrive in the main approximation to
\begin{equation}\label{Lam}
Q(E)=\Lambda(E)\,{b^{(s)}}^{\dag}b^{(s)}=\Lambda(E)
Q^{(s)}(E)\,;\quad \Lambda(E)=1+\frac{1}{2}\Gamma_s
l^{(e)}\;(E)
\end{equation}
where $l^{(e)}(E)$=$\frac{d}{\pi}$$\rm Tr$ $\left[{{\cal
G}_0^{(e)}}^{\dag}(E){\cal G}_0^{(e)}(E)\right]$.

A given bare doorway resonance state with the complex energy ${\cal
E}_0\!=\!\varepsilon_0\!-\!\frac{i}{2}\gamma_0$ generates a set of exact resonance
states with complex energies which are found from the eq.(\ref{Rs}). Since the energy
spectrum of the environment is very dense and rich, its states are supposed to decay
through a large number of weak statistically equivalent channels. The corresponding
decay amplitudes $A_{\mu}^{(e)}$ are random Gaussaian variables with zero means and
variances $\langle A_{\mu}^{(e)}A_{\nu}^{(e')}\rangle\!=\!\delta^{ee'}
\delta_{\mu\nu}\gamma_e/M^{(e)}$. The widths $\gamma_e$ does not fluctuate when
$M^{(e)}\gg$1. The interaction $V$ redistributes the initial widths over exact
resonances as $\Gamma_{\alpha}\!=\!f_{\alpha}\gamma_0\!+ \!(1-f_{\alpha})\gamma_e$
\cite{Sokolov1997} with the strength function $f_{\alpha}$$\equiv$ $f(E_{\alpha})$
obeying the condition $\sum_{\alpha}f_{\alpha}$=1.

Neglecting unobservable spectral fluctuations of the background we assume a rigid
spectrum with equidistant levels $\epsilon_{\mu}\!=\!\varepsilon_0\!+\!\mu d\!-\!
\frac{i}{2}\gamma_e$. One of the advantages of this uniform model
\cite{BohrMot1969,Sokolov1997} is that the loop functions $g^{(e)}(E), l^{(e)}(E)$ as
well as the strength function $f_{\alpha}$ can be calculated explicitly
\cite{Sokolov1997}. Depending on the magnitude of the coupling there exists two
different scenarios. In the limit $\Gamma_s\!\gg\!\gamma_0-\gamma_e$ (the natural
assumption $\gamma_e\!\ll\!\gamma_0$ is accepted throughout the paper) of strong
interaction all the individual strengths $f_{\alpha}$ are small,
$f(E_{\alpha})\!\leq\!2d/\pi\Gamma_s$, and are distributed around the energy
$\varepsilon_0$ according to the Lorentzian with the width $\Gamma_s$,
$f_{\alpha}\!\propto\!{\cal L}_{\Gamma_s}(E_{\alpha}\!- \!\varepsilon_0)$. Thus the
original doorway state fully dissolves in the sea of the background states. In the
opposite limit of weak coupling, $\Gamma_s\!\ll\!\gamma_0\!- \!\gamma_e$, which is the
one of our interest, the strength $f_0\!=\!1\!-\!\Gamma_s/(\gamma_0-\gamma_e)$ remains
large when the rest of them are small again
$f(E_{\alpha})\!\leq\!2d\,\Gamma_s/\pi(\gamma_0\!-\!\gamma_e)^2$ and distributed as
${\cal L}_{(\gamma_0\!-\!\gamma_e)}(E_{\alpha}-\varepsilon_0)$. Therefore, only in the
weak-coupling case the doorway state preserves individuality and can be observed through
the outer channels. The interaction with the background surrounds a doorway resonance by
a bunch of fine-structure resonances with Lorentzianly distributed heights.

3. When the scattering energy $E$ approaches the
environment ground energy only low-lying bare background
states are involved which are stable, $A_{\mu}^{(e)}\equiv
0$. The self-energy matrix $\Sigma^{(s)}(E)$ is hermitian
and, as a consequence, the scattering matrix $S(E)$ is
unitary in this limit. Note that the averaging over the
coupling $V$ does not spoil the unitarity.
\begin{equation}\label{g,l}
g^{(e)}(E)=\cot\left(\pi\frac{E-\varepsilon_0}{d}\right)\,;
\qquad l^{(e)}(E)=\frac{\pi}{d}\sin^{-2}\left(\pi\frac{E-
\varepsilon_0}{d}\right)\,.
\end{equation}

If the structure is almost closed so that the considered resonance is isolated the
individual cross sections
\begin{equation}\label{Icrs}
\sigma^{ab}(E)=\frac{\gamma^{(a)}\gamma^{(b)}}{\left[E-
\frac{1}{2}\Gamma_s\cot\left(\pi E/d\right)\right]^2+
\frac{1}{4}\gamma_0^2}\,;\quad\gamma^{(a)}=|A_0^a|^2\,;
\quad \gamma_0=\Sigma_a\gamma^{(a)}
\end{equation}
as well as the Wigner time delay
\begin{equation}\label{Q1}
\tau_W(E)= {\rm tr}\;
Q(E)=\gamma_0\;\frac{1+(\pi\Gamma_s/2d) \sin^{-2}\left(\pi
E/d\right)} {\left[E- \frac{1}{2}\Gamma_s\cot\left(\pi
E/d\right)\right]^2+ \frac{1}{4}\gamma_0^2}
\end{equation}
(we have set $\varepsilon_0=0$) reveal strong fine-structure fluctuations on the scale
of the background level spacing $d$. In particular, the delay time fluctuates between
$\tau_W(E\!=\!\epsilon_{\mu})\!=\!(2\pi/d)(\gamma_0/ \Gamma_s)\!\sim\!1/\Gamma_{\mu}$ at
the points of the fine structure levels and a much smaller value $\tau_W(E\!\approx\!0
\!\neq\!\epsilon_{\mu})\!\approx\!(2\pi/d)(\Gamma_s/ \gamma_0)$ in between. In fact, a
particular fine-structure resonance cannot be resolved and only quantities averaged over
an energy interval $d\!\ll\!\delta E\!\ll\!\Gamma_s, D$ are observed. (Equivalently, one
can use instead ensemble averaging over the fine-structure resonances.) Actually, it is
enough because of periodicity to average over the interval $-\frac{d}{2}< E
<\frac{d}{2}$. The averaging which can easily be performed explicitly yields
\begin{equation}\label{avIcrs}
\overline{\sigma^{ab}}(E)=\left(1+
\frac{\Gamma_s}{\gamma_0}\right)\,\frac{\gamma_a\gamma_b}
{E^2+\frac{1}{4}(\gamma_0+\Gamma_s)^2}
\end{equation}
and
\begin{equation}\label{avQ1}
\overline{\tau_W}(E)=\frac{\gamma_0+\Gamma_s}{E^2
+\frac{1}{4}(\gamma_0+\Gamma_s)^2}+\frac{2\pi}{d}\;.
\end{equation}
The first terms in the both expressions  correspond to excitation of a damped resonance
with the total width $\Gamma_0=\gamma_0+\Gamma_s$ when the second account for the
particles reinjected from the environment after the typical time delay proportional to
the background level density.

Generally, however, resonances overlap and the scattering amplitudes can be presented as
a sum of a large number $N^{(s)}\gg 1$ of interfering resonance contributions each one
being proportional to the energy dependent product
\begin{equation}\label{Eden}
\frac{1}{{\cal D}_n(E){\cal D}^*_{n'}(E)}=\frac{1}{{\cal
E}_n-{\cal E}^*_{n'}}\left[\frac{1}{{\cal
D}_n(E)}-\frac{1}{{\cal D}^*_{n'}(E)}\right]
\end{equation}
where ${\cal D}_n(E)=E-{\cal E}_n-
\frac{1}{2}\Gamma_s\cot\left(\pi\frac{E-E_n}{d}\right)$ and
$n,n'=1, 2, ..., N^{(s)}$. Using the identities:
\begin{equation}\label{Iden}
\overline{\frac{1}{{\cal D}_n(E)}}=\frac{1}{E-{\cal
E}_n+\frac{i}{2}\Gamma_s}\,;\quad \frac{1}{{\cal E}_n-{\cal
E}^*_{n'}}= i\int_0^{\infty}dt\, e^{-i({\cal E}_n-{\cal
E}^*_{n'})\;t}
\end{equation}
we arrive finally at an incoherent sum $\overline{\sigma^{ab}}(E)=
\sigma^{ab}_1(E)+\sigma^{ab}_2(E)$ of the flow formed by the interfering damped doorway
resonances
\begin{equation}\label{Sig1}
\sigma^{ab}_1(E)=\Big|\left(A^{\dag}\frac{1}{E- {\cal
H}^{(s)}+\frac{i}{2}\;\Gamma_s}A\right)^{ab}\Big|^2
\end{equation}
and given by the integral over all times $t>0$ retarded flow of the particles reemitted
by the the environment
\begin{equation}\label{Sig2}
\sigma^{ab}_2(E)=\Gamma_s\;\int_0^{\infty}dt\,
\Big|\left(A^{\dag}\frac{e^{-i{\cal H}^{(s)}\;t}}{E-{\cal
H}^{(s)}+\frac{i}{2}\;\Gamma_s}A\right)^{ab}\Big|^2\,.
\end{equation}
The contribution (\ref{Sig1}) is identical to the result of the Efetov's
imaginary-potential model if we identify the strength of this potential as
$\alpha=\frac{\pi}{D}\;\Gamma_s$. The second contribution (\ref{Sig2}) compensates the
loss of the flow due to absorption described by the first one.

For higher scattering energies, one should take into account that excited background
states are not stable and can strongly overlap, $\gamma_e\gg d$, even with no coupling
to the doorway states. This smears out the fine-structure fluctuations thus reducing the
loop functions to $g^{(e)}(E)\!\Rightarrow\!-i$ and
$l^{(e)}(E)\!\Rightarrow\!2/\gamma_e$ \cite{Sokolov1997}. As opposed to the above
consideration, the unitarity of the scattering submatrix $S$ is broken. The structures
narrower than $\gamma_e$ are excluded and all particles delayed for a time larger than
$1/\gamma_e$ are irreversibly absorbed and lost from the outgoing flow.

4. It follows that there exist two different temporal characteristics of the scattering
process (\ref{T}). The first one is the decay rate $\gamma_0\!+\!\Gamma_s$ of the
doorway state once excited through the incoming channel. This rate is readily seen from
the scattering amplitude ${\cal
T}(E)\!=\!\gamma_0\!\left[E-\varepsilon_0+\frac{i}{2}\left
(\gamma_0+\Gamma_s\right)\right]^{-1}$ near an isolated doorway resonance. Since the
doorway state is not an eigenstate of the total effective Hamiltonian $\mbox {\boldmath
${\cal H}$}$, it fades exponentially out not only because the particle is emitted onto
the outer channels but also due to the internal transitions with formation of the exact
fine structure resonances over which the doorway state spreads. More than that, even if
the background is stable the internal decay remains exponential up to the Heisenberg
time $2\pi/d$. Typically, during the time $\tau_s\!=\!1/\Gamma_s$ the background absorbs
the particle. After this, it can evade via one of the $e$-channels or be after a while
emitted back into the dot and finally escape onto an outer channel. Since the particle
reemitted by the chaotic background carries no phase memory, the characteristic time
during which this memory is lost (dephasing time) coincides with $\tau_s$. All the
resonances, save the broad one with the width $\Gamma_0$=$ \gamma_0-\Gamma_s$, have
rather small widths and decay much slower. Just the resonances with the widths within
the interval $\gamma_e\!<\!\Gamma_{\alpha}\!\ll\!\Gamma_s$ contributes principally in
the Wigner time delay which shows how long the excited system still returns particles
into the observed channel. For example the delay time near the energy of an isolated
doorway resonance equals
\begin{equation}\label{Q1s}
\tau_W(E)=\frac{\gamma_0}{E^2+\frac{1}{4}\left(\gamma_0+
\Gamma_s\right)^2}\Lambda\,.
\end{equation}
This result differs by the enhance factor $\Lambda$=1+$\Gamma_s/\gamma_e$ (compare with
eq. (\ref{avQ1})) from what follows from the somewhat oversimplified consideration in
\cite{SavSomm2003} which implies that interaction with an environment yields only
absorption.

Turning to general consideration, we model as usual the unperturbed chaotic
single-particle motion using the random matrix theory. The observable channels are
considered below to be statistically equivalent and, to simplify the calculation, no
time-reversal symmetry is suggested. We suppose below that $d\!\ll\!\gamma_e\!\ll\!\min
\left(D,\Gamma_s\right)\!\lll\!1$ (the radius of the semicircle). If, on the contrary,
$\gamma_e\!\gg\!\Gamma_s$ the transitions into the background become equivalent to
irreversible decay similar by its properties to the decay into continuum. In such a
limit of full absorption the factor $\Lambda\rightarrow 1$ and the approach
 of ref. \cite{SavSomm2003} becomes valid.

It is readily seen that to account for the interaction with the background the
substitution $\varepsilon\!\Rightarrow\!\varepsilon\!-\!i\Gamma_s$ should be done while
calculating the two-point S-matrix correlation function $C(\varepsilon)=S(E)\otimes
S^{\dag}(E+\varepsilon)$. This immediately yields the connection
$C_V(t)\!=\!\exp(-\Gamma_s t)C_0(t)$ between the Fourier transforms with and without
interaction, the additional damping being as before caused by the spreading over the
fine-structure resonances.

The modification given in \cite{SavSomm2003} of the method proposed in
\cite{Sommers2001} allows us to calculate also the distribution ${\cal P}(q)\!=\!(\pi
M^{(s)})^{-1}$Im$\langle$tr$(q-Q-i0)^{-1}\rangle$ of proper delay times (eigenvalues of
the Smith matrix). The corresponding generating function is proportional to the ratio of
the determinants of two $2N^{(s)}\times 2N^{(s)}$ matrices with the following structure
(compare with \cite{SavSomm2003})
\begin{eqnarray}\label{A}
A(z)&=&-i\left(E-H^{(s)}\right)\sigma_3+\frac{1}{2}\left(AA^{\dag}+
\Gamma_s\right)-\frac{\Lambda}{z}\left(1+\sqrt{1-
\frac{\Gamma_s}{\Lambda}z}\;\sigma_1\right) \nonumber \\
&\approx&-i\left(E-H^{(s)}\right)\sigma_3+\frac{1}{2}AA^{\dag}-
\left(1+\sigma_1\right)\left(\frac{\Lambda}{z}-
\frac{\Gamma_s}{2}\right)
\end{eqnarray}
where the variable $z$ spans the complex $q$-plane. The square root sets
\cite{SavSomm2003} the restriction $q\leq\Lambda/\Gamma_s\!\approx\!1/\gamma_e$ on the
positive real axes. In the approximation of the second line valid if $\Lambda\!\gg\! 1$
we arrive to a simple relation
\begin{equation}\label{P(tau)}
{\cal P}_V(\tau)=\frac{1}{\Lambda}{\cal P}_0(\tau_V)\;,
\end{equation}
where $\tau_V\!\equiv\!\left(\tau/ \Lambda\right)\left[1-\pi\left(\Gamma_s/\Lambda
D\right) \tau\right]^{-1}\!>\!0$ and $\tau\!=\!\frac{D}{2\pi} q$. This relation does not
hold in the asymptotic region $\tau_V$$\rightarrow$$\infty$ or
$\tau$$\rightarrow$$\Lambda D/\pi\Gamma_s$$\approx$$\frac{1}{\pi}D/\gamma_e$ where the
more elaborate rigorous expressions obtained in \cite{SavSomm2003} must be used with the
substitution $\tau$$\Rightarrow$$\tau/\Lambda$ being made. For example in the
single-channel scattering with the perfect coupling to the continuum the time delay
distribution (see \cite{FyodSomm1997}) ${\cal P}_V(\tau)\!=\!e^{-1/\tau_V}/\tau_V^3$
reaches its maximum at the point $\tau_V\!=\!1/3$ or $\tau\!=\!(1/3)\Lambda\left(1+
\frac{\pi}{3}\Gamma_s/D\right)^{-1}\!\gg\!1/3$. The most probable delays shift towards
larger values. The estimation just made is quantitatively valid only if
$\Gamma_s\!\ll\!D/2\pi$ and the maximum lies near the point
$\frac{1}{3}\Gamma_s/\gamma_e$ distant from the exact edge $D/2\pi\gamma_e$ of the
distribution. Under the latter restriction the approximate formula (\ref{P(tau)})
describes well the bulk of the delay time distribution which becomes, roughly, $\Lambda$
times wider. The condition noted is much less restrictive in the case of weak coupling
to the continuum when the transmission coefficient $T\ll 1$ and the most probable delay
time $\tau\!\approx\!T\Gamma_s/4\gamma_e\! \ll\!D/2\pi\gamma_e$ as long as
$\Gamma_s\!\ll\!D/T\pi$.

The approximation (\ref{P(tau)}) works even better when the number of channels
$M^{(s)}\!\gg\!1$ and the delay times are restricted to a finite interval
\cite{Brouwer1997,Sommers2001}. The delay time scales in this case with ${M^{(s)}}^{-1}$
and the natural variable looks as $\tau\!=\!qM^{(s)}D/2\pi\!=\!\Gamma_W q/T$ with
$\Gamma_W$ being the Weisskopf width. The edges of the distribution (\ref{P(tau)}) are
displace towards larger delays by the factor $\Lambda$,
$\tau_V^{(\mp)}\!=\!\Lambda\tau_0^{(\mp)}$. One can readily convince oneself that the
taken approximation remains valid in the whole interval
$\Delta\tau_V\!=\!\Lambda(\tau_0^{(+)}\!-\!\tau_0^{(-)})$ under condition
$\Gamma_s\!\ll\!\Gamma_W$ which implies weak interaction with the background. The width
of the delay time distribution broadens by the factor $\Lambda$ due to the influence of
the background.

5. In summary, the influence of a complicated environment on the chaotic single-particle
scattering is analyzed. Unlike some earlier considerations, the coupling to the
background is supposed to be purely Hermitian. The single-particle doorway states which
are excited inside a mesoscopic device and observed through external channels are
additionally damped with the rate $\Gamma_s\!=\! 2\pi\langle |V|^2\rangle/d$ because of
the spreading over the long-lived fine-structure resonances. This rate uniquely
determines the dephasing time during the particle transport through the ballistic
microstructure. Absorption takes place because of hidden decays of the background
resonance states. The formation of the fine-structure resonances strongly enhances delay
time fluctuations thus, in particular, broadening the distribution of the proper delay
times.

I am grateful to Y.V. Fyodorov and D.V. Savin for useful discussions and critical
remarks. The financial support from RFBR through Grant No 03-02-16151 is acknowledged
with thanks.

\end{document}